\documentclass[12pt]{article}
\usepackage[usenames,dvipsnames,svgnames,table]{xcolor} 
\usepackage{kotex}
\usepackage{graphicx}
\usepackage{epsfig}
\usepackage{amssymb, amsmath, mathtools}
\usepackage{verbatim}
\usepackage{natbib}
\usepackage[affil-it]{authblk}
\usepackage{mathrsfs}
\usepackage{here}
\usepackage{makecell}
\usepackage{tikz}
\usepackage{enumerate}
\usepackage[shortlabels]{enumitem}
\usepackage{yfonts}
\usepackage{comment}
\usepackage{longtable}
\usepackage{tikz,amsmath, amssymb,bm,color}
\usetikzlibrary{circuits.logic.US,circuits.logic.IEC,fit}
\usepackage{algorithm2e}
\usepackage{multirow}
\usepackage{physics}
\usepackage{amsthm}
\usepackage{makecell}
\usepackage{ulem}
\usepackage{proofread}
\usepackage{booktabs}
\usepackage{array}
\usepackage{xr}
\usepackage{adjustbox}
\usepackage{lineno}
\usepackage{import}

\usepackage[colorlinks]{hyperref} 
\usepackage[colorinlistoftodos, textsize=scriptsize]{todonotes} 

\usepackage[framemethod=TikZ]{mdframed}
\mdfdefinestyle{JL}{%
	linecolor=blue,
	outerlinewidth=1pt,
	roundcorner=2pt,
	innertopmargin=0.1\baselineskip,
	innerbottommargin=0.1\baselineskip,
	innerrightmargin=2pt,
	innerleftmargin=2pt,
	backgroundcolor=orange!100!white}

\hypersetup{
	colorlinks,
	citecolor=Black,
	linkcolor=Red,
	urlcolor=Blue}

\includecomment{note} 

\newtheorem{theorem}{Theorem}[section]
\newtheorem{lemma}[theorem]{Lemma}
\newtheorem{definition}[theorem]{Definition}
\newtheorem{proposition}[theorem]{Proposition}
\newtheorem{corollary}[theorem]{Corollary}

\newtheorem{remark}[theorem]{Remark}

\everymath{\displaystyle}

\newcommand{\be}{\begin{equation}}
	\newcommand{\ee}{\end{equation}}
\newcommand{\bea}{\begin{eqnarray*}}
	\newcommand{\eea}{\end{eqnarray*}}
\newcommand{\bean}{\begin{eqnarray}}
	\newcommand{\eean}{\end{eqnarray}}
\newcommand{\ben}{\begin{enumerate}}
	\newcommand{\een}{\end{enumerate}}
\newcommand{\bi}{\begin{itemize}}
	\newcommand{\ei}{\end{itemize}}
\newcommand{\brem}{\begin{remark}}
	\newcommand{\erem}{\end{remark}}
\newcommand{\bcen}{\begin{center}}
	\newcommand{\ecen}{\end{center}}
\newcommand{\bsv}{\begin{semiverbatim}}
	\newcommand{\esv}{\end{semiverbatim}}

\newcommand{\bt}{\begin{theorem}}
	\newcommand{\et}{\end{theorem}}
\newcommand{\bl}{\begin{lemma}}
	\newcommand{\el}{\end{lemma}}
\newcommand{\bd}{\begin{definition}}
	\newcommand{\ed}{\end{definition}}
\newcommand{\bc}{\begin{corollary}}
	\newcommand{\ec}{\end{corollary}}
\newcommand{\bp}{\begin{proposition}}
	\newcommand{\ep}{\end{proposition}}

\newcommand{\bfX}{ \mathbf{X}}

\newcommand{\bbE}{{ \mathbb{E}}}

\newcommand{\bbR}{ \mathbb{R}}

\usepackage{xr}
\makeatletter
\newcommand*{\addFileDependency}[1]{
	\typeout{(#1)}
	\@addtofilelist{#1}
	\IfFileExists{#1}{}{\typeout{No file #1.}}
}
\makeatother

\modulolinenumbers[5]

\title{Bayesian Bootstrap based Gaussian Copula Model for Mixed Data with High Missing Rates}
\author[1]{Seongmin Kim}
\author[1]{Jeunghun Oh}
\author[2]{Hungkuk Ko}
\author[2]{Jeongmin Park}
\author[1]{Jaeyong Lee}
\affil[1]{Department of Statistics, Seoul National University}
\affil[2]{Manufacturing Quality and Reliability Group, Samsung Electronics}

\allowdisplaybreaks
\begin{document}
	
\maketitle

\begin{abstract}
Missing data is a common issue in various fields such as medicine, social sciences, and natural sciences, and it poses significant challenges for accurate statistical analysis. Although numerous imputation methods have been proposed to address this issue, many of them fail to adequately capture the complex dependency structure among variables. To overcome this limitation, models based on the Gaussian copula framework have been introduced. However, most existing copula-based approaches do not account for the uncertainty in the marginal distributions, which can lead to biased marginal estimates and degraded performance, especially under high missingness rates.

In this study, we propose a Bayesian bootstrap-based Gaussian Copula model (BBGC) that explicitly incorporates uncertainty in the marginal distributions of each variable. The proposed BBGC combines the flexible dependency modeling capability of the Gaussian copula with the Bayesian uncertainty quantification of marginal cumulative distribution functions (CDFs) via the Bayesian bootstrap. Furthermore, it is extended to handle mixed data types by incorporating methods for ordinal variable modeling.

Through simulation studies and experiments on real-world datasets from the UCI repository, we demonstrate that the proposed BBGC outperforms existing imputation methods across various missing rates and mechanisms (MCAR, MAR). Additionally, the proposed model shows superior performance on real semiconductor manufacturing process data compared to conventional imputation approaches.
\end{abstract}

\section{Introduction} 
Missing data frequently arises across a wide range of fields, including medicine, natural sciences, and social sciences. In industrial process data, where massive amounts of data are collected and managed, missingness can occur due to various factors such as sensor failures, routine maintenance, improper sampling at scheduled times, discarding of data with gross measurement errors, heterogeneous sampling rates across sensors, or temporary sensor shutdowns for maintenance (\citealt{arteaga2002dealing}; \citealt{muteki2005estimation}; \citealt{nelson1996missing}). In clinical data, missingness may result from unrecorded participant information, difficult-to-measure laboratory tests or biomarkers, or high dropout rates in longitudinal studies (\citealt{ibrahim2012missing}). For DNA microarray data, missing values often arise from physical issues such as low resolution, image distortion, dust or scratches on the slide, or problems related to spot quality control (\citealt{troyanskaya2001missing}; \citealt{aittokallio2010dealing}).

There are two main approaches to handling missing data. The first is to ignore the missing values by removing the observations or variables that contain them, thereby producing a complete dataset. According to \cite{acuna2004treatment}, deletion strategies (i.e., ignoring data) can be effective when the proportion of missing values is below 15\%. However, when applied to datasets with a missing rate above 15\%, these methods tend to yield biased results. In fact, \cite{janssen2010missing} showed that imputation methods can reduce bias in statistical analyses compared to deletion-based approaches. As a result, instead of ignoring missing data, researchers have proposed various imputation methods that fill in missing values to generate a complete dataset. Early imputation methods often relied on simple techniques such as mean imputation or regression-based approaches. While these methods are computationally efficient, they tend to underestimate variability in the data and may introduce bias in subsequent analyses. More sophisticated approaches, such as the Expectation-Maximization (EM) algorithm, iteratively update parameter estimates based on imputed values and offer more statistically sound results (\citealt{little2019statistical}).

\cite{rubin1976inference} and \cite{little2019statistical} proposed a classification of missing data mechanisms into three types: missing completely at random (MCAR), missing at random (MAR), and missing not at random (MNAR). When the probability of missingness is independent of both observed and unobserved values, the data are said to be MCAR. If the missingness depends only on the observed values but not on the missing values, the mechanism is considered MAR. In contrast, when the missingness depends on the unobserved values themselves, it is referred to as MNAR. 

Identifying the missing data mechanism is crucial when applying imputation methods in real-world data analysis. For example, applying mean imputation to data where the missingness follows a MAR mechanism can lead to biased imputations and incorrect statistical inferences. The MNAR assumption is particularly challenging, as the unobserved values are themselves related to the cause of missingness, making imputation based solely on the observed data difficult or invalid. As a result, many imputation methods proposed in the literature assume MAR. In this study, we also focus on imputation under the MAR mechanism.

A variety of imputation methods have been proposed to handle missing data. One of the most widely used approaches is Multiple Imputation by Chained Equations (MICE), which offers the advantage of capturing the uncertainty inherent in the missing data through multiple imputations (\citealt{azur2011multiple}; \citealt{van2011mice}). In addition, nonparametric methods such as $k$-nearest neighbors (KNN) imputation have been introduced, which estimate missing values based on the similarity between observed samples (\citealt{troyanskaya2001missing}). The missForest method, a random forest-based imputation technique, utilizes ensemble learning techniques to model complex interactions among variables and to predicts missing values (\citealt{stekhoven2012missforest}). Other approaches based on matrix factorization and low-rank approximation have also been explored, particularly for recovering the underlying structure of high-dimensional data (\citealt{hastie1999imputing}). Recently, deep learning-based methods such as autoencoder imputation have gained attention for their ability to capture nonlinear patterns in the data and deliver promising results (\citealt{gondara2018mida}). In addition, Bayesian nonparametric approaches such as Dirichlet Process Mixture Models (DPMM) have been applied to address missingness in semiconductor manufacturing data (\citealt{park2023bayesian}).

Copula-based imputation techniques have gained considerable attention in addressing multivariate missing data due to their ability to separately and flexibly model marginal distributions and dependence structures. This flexibility enables the capture of complex inter-variable dependencies. Early copula-based approaches relied on empirical cumulative distribution functions (ECDFs), whereas later methods adopted rank likelihood to avoid the direct estimation of marginal distributions (\citealt{hoff2007extending}; \citealt{mulgrave2020bayesian}).

More recently, \cite{mulgrave2023bayesian} proposed a Bayesian model that places priors on marginal cumulative distribution functions using B-spline basis functions, allowing for posterior inference over the marginal distributions. \cite{zhao2020missing} and \cite{feldman2024nonparametric} applied Gaussian copula models for imputing missing values in mixed-type data, while \cite{murray2013bayesian} introduced a Gaussian copula factor model for modeling structured latent dependencies. Among these, the models of \cite{mulgrave2023bayesian} and \cite{feldman2024nonparametric} enable uncertainty quantification by adopting fully Bayesian approaches. However, despite this flexibility, they do not provide explicit theoretical guarantees on the convergence rates of the marginal distributions.

In this paper, we propose the Bayesian bootstrap-based Gaussian Copula (BBGC) model, a novel, theoretically grounded framework for multivariate missing data imputation. BBGC enables uncertainty quantification over marginal distributions in a computationally efficient manner. It integrates the Bayesian bootstrap, originally introduced by \cite{rubin1981bayesian}, into a Gaussian copula framework, providing a fully nonparametric alternative to ECDF or spline-based marginal modeling. Our method offers a quantifiable finite-sample convergence guarantee for the marginal distribution functions derived via the Bayesian bootstrap. This renders our approach both practically efficient and theoretically grounded. Furthermore, by incorporating the ordinal transformation technique proposed by \cite{zhao2020missing}, BBGC is extended to accommodate mixed-type data, including both continuous and ordinal variables.

The remainder of this paper is organized as follows. Section~\ref{sec:preli} provides a review of the Gaussian copula and the Bayesian bootstrap, and introduces notation used throughout the paper. Section~\ref{sec:model} describes the proposed model with theoretical result of the Bayesian bootstrap and outlines the corresponding sampling procedure. Section~\ref{sec:simul_studies} presents simulation studies under various missingness patterns, and Section~\ref{sec:real_data} reports empirical results on real datasets. Finally, Section~\ref{sec:conclusion} concludes the paper.

\section{Preliminaries}\label{sec:preli}
\subsection{Notation}
Let $\bfX = (x_{ij})$ be an $n \times p$ data matrix, where $n$ denotes the number of observations and $p$ denotes the number of variables. Each entry $x_{ij}$ may be continuous, ordinal, or missing. If the $j$-th variable is continuous, then $x_{ij} \in \mathbb{R}$. If the $j$-th variable is ordinal, then $x_{ij} \in \{1, \ldots, l_j \}$, where $l_j$ denotes the number of ordinal categories for the $j$-th variable.

We define $\mathcal{O}_c := \{(i, j) : x_{ij} \text{ is observed, and the $j$-th variable is continuous} \}$ and $\mathcal{O}_o := \{(i, j) : x_{ij} \text{ is observed and the $j$-th variable is ordinal} \}$ denote the sets of observed entries for continuous and ordinal variables, respectively. The corresponding sets of missing entries are denoted by $\mathcal{M}_c$ and $\mathcal{M}_o$ for continuous and ordinal variables, respectively. Let $\bfX_{\mathcal{O}}$ and $\bfX_{\mathcal{M}}$ denote the observed and missing data, respectively, where
$$\bfX_{\mathcal{O}} = \{x_{ij}:(i,j)\in\mathcal{O}_c\cup\mathcal{O}_o\},\quad \bfX_{\mathcal{M}} = \{x_{ij}:(i,j)\in\mathcal{M}_c\cup\mathcal{M}_o\}.$$

We write $(w_1,\ldots,w_n) \sim \text{Dir}(\alpha)$ for $\alpha = (\alpha_1,\ldots,\alpha_n)$ with $\alpha_i > 0$ for $i = 1,\ldots,n$. Its density is proportional to $\prod_{i=1}^n w_i^{\alpha_i - 1}$ on the simplex $\{ (w_1, \ldots, w_n) \in [0,1]^n : \sum_{i=1}^n w_i = 1 \}$. We write $\Sigma \sim \text{Inv-Wishart}(\nu, \Psi)$ where $\nu > p - 1$ and $\Psi$ is a $p \times p$ positive definite matrix. The density of $\Sigma$ is proportional to $\abs{\Sigma}^{-(\nu+p+1)/2} \exp\big( -\dfrac{1}{2}tr(\Psi \Sigma^{-1})\big)$,
where $\Sigma$ is a $p \times p$ positive definite matrix.

For $R\in\bbR^{p\times p}$, $R_{j,-j}$ denotes the $j$-th row of $R$ with the $j$-th element excluded, yielding a $1\times(p-1)$ row vector. Likewise, $R_{-j,-j}$ denotes the submatrix obtained by removing the $j$-th row and column from $R$, resulting in a $(p-1)\times (p-1)$ matrix. The notation $R_{-j,j}$ corresponds to the $j$-th column of $R$ with the $j$-th entry removed, and is equivalent to the transpose of $R_{j,-j}$.

Let $\Phi(\cdot)$ denote the cumulative distribution function (CDF) of the standard normal distribution, and let $\Phi_R(\cdot)$ denote the multivariate normal CDF with mean vector $\mathbf{0}_{p \times 1}$ and covariance matrix $R$.
We also define the Dirac measure $\delta_x$, which satisfies $\delta_x(A) = 1$ if $x \in A$ and $\delta_x(A) = 0$ if $x \notin A$ for all set $A \subset \mathbb{R}$.
For a real number $x \in \mathbb{R}$, let $\lfloor x \rceil$ denote the nearest integer to $x$; that is, the rounding function that returns the integer closest to $x$.

\subsection{Gaussian Copula Model}
Copula functions enables the modeling of multivariate distributions by separating the marginal distributions from the dependence structure. Specifically, let $X = (X_1, \ldots, X_p)^T$ be a $p$-dimensional random vector with joint cumulative distribution function (CDF) $F_X(x_1,\ldots,x_p)$ and marginal CDFs $F_i(x_i)$ for each variable $X_i$. Then there exists a function $C: [0, 1]^p \rightarrow [0, 1]$ such that (\citealt{sklar1959fonctions})
\begin{equation*}
F_X(x_1,\ldots,x_p) = C\bigl(F_1(x_1), \ldots, F_p(x_p)\bigr),
\end{equation*}
where $C$, known as the copula function, captures the dependence structure among the variables.

The Gaussian copula model represents the dependence structure using a multivariate normal distribution. The Gaussian copula function is given by
\begin{equation*}
C(u_1,\ldots,u_p) = \Phi_R\Bigl(\Phi^{-1}(u_1), \ldots, \Phi^{-1}(u_p)\Bigr),
\end{equation*}
where $u_i \in [0,1]$ for $i = 1, \ldots, p$. 

For a random vector $X = (X_1, \ldots, X_p)^T$, the following transformation holds:
\begin{equation} \label{form:varTransform}
\Phi^{-1}\bigl(F_i(X_i)\bigr) \sim N(0,1), \quad \text{for } i = 1, \ldots, p.
\end{equation}
Assuming the transformed variables jointly follow a multivariate normal distribution, we obtain:
\begin{equation*}
\Bigl(\Phi^{-1}(F_1(X_1)), \ldots, \Phi^{-1}(F_p(X_p))\Bigr)^T \sim N(0, R),
\end{equation*}
where $R$ is a $p \times p$ correlation matrix. This structure allows for flexible modeling of dependencies among variables via the multivariate normal assumption.

\subsection{Bayesian Bootstrap}
Standard bootstrap methods generate samples via simple random sampling with replacement. In contrast, \cite{rubin1981bayesian} proposed the Bayesian bootstrap, which models the distribution of the data using a Dirichlet distribution. Given observations $X_1, \ldots, X_n$, and their empirical distribution function $F_n$ defined as $F_n(t) = \sum_{i=1}^n \dfrac{1}{n} I(X_i \leq t)$, the Bayesian bootstrap cumulative distribution function $F^{BB}$ is defined as follows:
\begin{align*} 
F^{BB}(t) &= \sum_{i=1}^n w_i I(X_i\leq t),\\
(w_1, \ldots, w_n) &\sim  Dir(1, \ldots, 1),
\end{align*}
where the weights $(w_1, \ldots, w_n)$ satisfy $\sum_{i = 1}^n w_i = 1$. When $w_i = 1/n$ for $ i = 1, \ldots,n$, the Bayesian bootstrap CDF coincides with the empirical distribution function $F_n$.

By assigning a probabilistic model to the weights of the empirical CDF, the Bayesian bootstrap accounts for uncertainty in the cumulative distribution. This approach is particularly useful in statistical analysis involving small sample sizes or complex data structures, where the Bayesian bootstrap is known to provide more reliable uncertainty quantification than standard bootstrap methods (\cite{mostofian2019statistical}).

\section{Bayesian Bootstrap-based Gaussian Copula}\label{sec:model}
\subsection{Posterior under MAR}
Traditional imputation models often assume that data is missing completely at random (MCAR), but in practice, the more flexible missing at random (MAR) assumption is more realistic. To establish a principled foundation for posterior inference under the MAR setting, we begin by presenting the following result.

\begin{lemma}\label{lem:posterior_mar}
Consider the Gaussian copula model:
\begin{equation*}
\Bigl(\Phi^{-1}(F_1(X_1)), \ldots, \Phi^{-1}(F_p(X_p))\Bigr)^T \sim N(0, R),
\end{equation*}
where $R$ is a $p\times p$ correlation matrix. Let $\mathcal{M} = \mathcal{M}_c\cup \mathcal{M}_o$ denote the missingness indicator, and let $F = (F_1, \ldots, F_p)$ denote the marginal CDFs. Under the MAR assumption, the posterior distribution satisfies
\begin{align*}
\pi(R \vert \bfX_{\mathcal{O}}) 
&= \int \pi(R \vert \bfX_{\mathcal{O}}, F) \cdot \pi(F \vert \bfX_{\mathcal{O}}) \, dF.
\end{align*}
\end{lemma}

This result follows from two key conditions. First, the MAR assumption allows us to ignore the missingness mechanism $P(\mathcal{M} \vert \bfX)$ in the posterior distribution. Second, we assume a conditional independence structure in which the posterior distribution of $R$ depends on the full data only through its transformation via the marginal CDFs $F$. These two conditions together enable marginalization over the missing values and lead to a tractable posterior for $R$ based solely on the observed data and the transformation $F$. This formulation provides the theoretical foundation for the model introduced in the next section.

\subsection{Model}
Based on Lemma~\ref{lem:posterior_mar}, we propose a Bayesian bootstrap-based Gaussian copula (BBGC) model that integrates the uncertainty of the marginal cumulative distribution functions via Bayesian model averaging. 

Traditional Gaussian copula models either use point estimates of the marginal distributions via the empirical CDF or bypass them using extended rank likelihood. These approaches typically do not account for uncertainty in marginal CDFs. In contrast, the BBGC model captures this uncertainty by sampling marginal CDFs from the Bayesian bootstrap posterior. Additionally, by incorporating the cutoff transformation in \cite{zhao2020missing}, the model accommodates both continuous and ordinal variables.

Formally, the marginal posterior of the correlation matrix $R$ is given by:
\begin{align*}
    \pi(R \vert \bfX_{\mathcal{O}}) &\propto \int \pi(R \vert \bfX_{\mathcal{O}}, F) \cdot \pi(F \vert \bfX_{\mathcal{O}}) \, dF,
\end{align*}
where $\bfX_{\mathcal{O}}$ denotes the observed data and $F = (F_1, \ldots, F_p)$ are the marginal CDFs.

To incorporate the uncertainty of the marginal distribution functions $F$ within the model, we adopt a Bayesian model averaging approach, in which we average over multiple realizations of $F$ drawn from the distribution $\pi(F \vert \bfX_{\mathcal{O}})$. The details of the distribution $\pi(F \vert \bfX_{\mathcal{O}})$ are described in Subsection \ref{subsec:adj_bb}.

For a given $F$, the conditional posterior becomes:
\begin{align}\label{eq:cond_post}
\pi(R \vert \bfX_{\mathcal{O}}, F) &\propto \pi(R) \cdot \int P(\bfX_{\mathcal{O}} \vert Z, F) \cdot P(Z \vert R) \, dZ,
\end{align}
where $Z \in \mathbb{R}^{n \times p}$ are latent Gaussian variables. The likelihood term is:
\begin{align*}
P(\bfX_{\mathcal{O}} \vert Z, F) =
\prod_{(i,j) \in \mathcal{O}_c}
\delta_{z_{ij}}( \{\Phi^{-1}(F_j(x_{ij}))\}) \cdot 
\prod_{(i,j) \in \mathcal{O}_o}
\delta_{z_{ij}}((s_{j,x_{ij}-1}, s_{j,x_{ij}}]),
\end{align*}
with cutoff values $s_{j,\ell}$ derived from $\tilde{F}_j^{BB}$. The detailed formulation of the likelihood term $P(\bfX_{\mathcal{O}} \vert Z, F)$ is provided in Subsection~\ref{subsec:marg_trans}.

This transformation ensures that each latent variable has standard normal margins, and the dependency structure is captured by the Gaussian copula:
\begin{align*}
Z_i \vert R \overset{iid}{\sim} \mathcal{N}(0, R).
\end{align*}

The prior distribution $\pi(R)$ is specified using a restricted inverse-Wishart construction to ensure that $R$ is a valid correlation matrix—i.e., symmetric, positive definite, and with unit diagonal entries. Specifically, a covariance matrix $R^*$ is sampled as:
\begin{align*}
R^* \sim \mathrm{Inv\text{-}Wishart}(\nu_0, \Psi_0),\quad \nu_0 > p-1,
\end{align*}
where $\Psi_0$ is a $p\times p$ positive definite matrix, and then normalized to obtain a valid correlation matrix:
\begin{align*}
R = \text{diag}(R^*)^{-1/2} R^* \, \text{diag}(R^*)^{-1/2}.
\end{align*}
In Subsection \ref{subsec:gibbs}, we describe the Gibbs sampling procedure in detail for posterior inference.

\subsection{Adjusted Bayesian bootstrap}\label{subsec:adj_bb}
Let $F_j^{BB}$ denote the Bayesian bootstrap CDF obtained from variable $X_j$. Directly using this function in the transformation \eqref{form:varTransform} leads to two main issues.

First, consider the minimum and maximum values of variable $j$ in the data, denoted as $x_{\min,j} := \min \{x_{ij} : i = 1, \ldots, n \}$ and $x_{\max,j} := \max \{x_{ij} : i = 1, \ldots, n \}$. Since $F_j^{BB}(x_{\max,j}) = 1$, applying the inverse CDF yields divergence:
\begin{align*}
\Phi^{-1} (F_j^{BB} (x_{\max,j}) ) = \Phi^{-1} (1) = \infty.
\end{align*}

Second, this transformation violates sign invariance, which requires that the transformation be invariant under negation of the variable, i.e., the model using $X$ or $-X$ should yield the same result. This condition can be stated as:
\begin{align*}
F_j (x_{ij}) = 1 - F_j^* (-x_{ij}),
\end{align*}
where $F_j$ and $F_j^*$ are the CDFs derived from $X_j$ and $-X_j$, respectively. The Bayesian bootstrap CDF $F_j^{BB}$ does not satisfy the symmetry condition at the boundaries
\begin{align*}
\mathbb{E}[F_j^{BB}(x_{\min,j})] \neq 1 - \mathbb{E}[F_j^{BB}(x_{\max,j})].
\end{align*}

\begin{theorem}\label{thm:sign_inv_cdf}
    Assume that $X_1, \ldots, X_n \overset{iid}{\sim} F$, where $F$ is a continuous distribution function. Let $F^{BB}$ denote the Bayesian bootstrap CDF defined by
    $$
    F^{BB}(t) = \sum_{i=1}^n w_i I(X_i \leq t), \quad (w_1, \ldots, w_n) \sim \mathrm{Dir}(1, \ldots, 1).
    $$
    Consider a linear transformation of $F^{BB}(t)$ given by
    $$\tilde{F}^{BB}(t) = a F^{BB}(t) + b,\quad a, b \in \mathbb{R}.$$
    If the coefficients satisfy the condition
    $$\dfrac{n+1}{n}a + 2b = 1,$$
    then $\tilde{F}^{\mathrm{BB}}(t)$ is sign-invariant.
\end{theorem}

To address both the divergence of the inverse CDF and the issue of sign invariance, we adopt $a = \dfrac{n}{n+1}$ and $b = 0$ in Theorem~\ref{thm:sign_inv_cdf}, which is equivalent to the adjustment proposed in \cite{madsen2006methods}. The resulting adjusted Bayesian bootstrap CDF $\tilde{F}^{BB}$ is given by
\begin{align*}
    \tilde{F}_j^{BB}(t) = \frac{n}{n+1} F_j^{BB}(t).
\end{align*}

This adjustment ensures $\tilde{F}_j^{BB}(x_{\max,j}) < 1$ and restores sign invariance.

\begin{theorem}[Posterior convergence rate of Bayesian bootstrap CDF]\label{thm:post_conv_cdf}
Assume that $X_i\overset{iid}{\sim} F$ for $i=1,\ldots,n$, where $F$ is a continuous distribution function. Let $\tilde{F}^{BB}$  denote the adjusted Bayesian bootstrap CDF given by
\begin{align*}
\tilde{F}^{BB}(t) = \dfrac{n}{n+1}\sum_{i=1}^n w_i I(X_i\leq t), \quad (w_1,\ldots,w_n) \sim  Dir(1, \ldots, 1).
\end{align*}
Then
$$\pi(\sup_{t\in\bbR}\abs{\tilde{F}^{BB}(t)-F(t)} >\delta  \vert \bfX ) \leq 4\exp\Big(-\dfrac{1}{2}(n-1)(\delta - \dfrac{2n}{n^2-1})^2 \Big),$$
for all $\delta>\dfrac{2n}{n^2-1}$ and for all integer $n\geq 2$.
\end{theorem}

Theorem \ref{thm:post_conv_cdf} shows that the adjusted Bayesian bootstrap CDF converges uniformly to the true distribution $F$, with a convergence rate that is of the same order as that of the empirical CDF. \cite{lo1987large} established that the Bayesian bootstrap CDF converges asymptotically to the empirical CDF. In contrast, Theorem \ref{thm:post_conv_cdf} provides a finite-sample guarantee. These results support the Bayesian bootstrap CDF as a natural and reliable extension of the empirical distribution, even in small sample settings.

\subsection{Latent Variable Transformation}\label{subsec:marg_trans}

We now describe the transformation of the observed data $\bfX_{\mathcal{O}}$ into the latent Gaussian variables $Z$, using the adjusted Bayesian bootstrap CDFs $\tilde{F}_j^{BB}$.

For continuous entries $(i,j) \in \mathcal{O}_c$, we define:
$$
z_{ij} = \Phi^{-1}(\tilde{F}_j^{BB}(x_{ij})).
$$

For ordinal entries $(i,j) \in \mathcal{O}_o$ where $x_{ij} \in \{1, \ldots, l_j\}$, the variable is mapped to a continuous latent value via thresholding
$$
z_{ij} \sim U(s_{j, x_{ij}-1}, s_{j, x_{ij}}], \quad
\text{where } s_{j, \ell} = \tilde{F}_j^{BB}(\ell), \quad \ell = 1, \ldots, l_j - 1,
$$
with boundaries set as $s_{j,0} = -\infty$ and $s_{j, l_j} = \infty$.

Finally, the likelihood for all observed variables, given $Z$ and $F$, is:
\begin{align*}
P(\bfX_{\mathcal{O}} \vert Z, F) =
\prod_{(i,j) \in \mathcal{O}_c}
\delta_{z_{ij}}( \{\Phi^{-1}(F_j(x_{ij}))\}) \cdot 
\prod_{(i,j) \in \mathcal{O}_o}
\delta_{z_{ij}}((s_{j,x_{ij}-1} , s_{j,x_{ij}}]),
\end{align*}
ensuring compatibility between observed data and the latent structure induced by the Gaussian copula.

\subsection{Gibbs Sampling Procedure}\label{subsec:gibbs}
Inference in BBGC is performed via Gibbs sampling, where the latent variable matrix $\mathbf{Z}$ and the correlation matrix $R$ are alternately sampled. Each iteration draws from the conditional distribution, conditioned on the marginal transformation $F$. 

According to equation \eqref{eq:cond_post}, the joint posterior distribution satisfies:
\begin{align*}
\pi(R, Z \vert \bfX_{\mathcal{O}}, F) &\propto \pi(R) \cdot  P(\bfX_{\mathcal{O}} \vert Z, F) \cdot P(Z \vert R)\\
&= \pi(R) \cdot  P(\bfX_{\mathcal{O}} \vert Z_{\mathcal{O}}, F) \cdot P(Z \vert R)
\end{align*}
where $Z_{\mathcal{O}} = \{Z_{ij}: (i,j)\in\mathcal{O}_c\cup\mathcal{O}_o\}$ represents the latent variables associated with the observed entries. The sampling procedure for $\pi(R,Z\vert \bfX_{\mathcal{O}}, F)$ consists of three main steps:

\begin{itemize}
  \item \textbf{Step A: Sample latent variables $Z_{ij}$}
  
  For each $(i,j)$, the update rule depends on both the variable type and its observation status:
  \begin{itemize}
    \item If $(i,j) \in \mathcal{O}_c$ (observed continuous):
    $$Z_{ij} = \Phi^{-1}(F_j(x_{ij})).$$
    
    \item If $(i,j) \in \mathcal{M}_c$ (missing continuous):
    $$Z_{ij} \sim \mathcal{N}(\mu_{ij}, \sigma_{ij}^2).$$

    \item If $(i,j) \in \mathcal{O}_o$ and $x_{ij} = \ell$ (observed ordinal):
    $$Z_{ij} \sim \mathcal{N}(\mu_{ij}, \sigma_{ij}^2) 
    \quad \text{truncated to } (s_{j,\ell-1}, s_{j,\ell}].$$

    \item If $(i,j) \in \mathcal{M}_o$ (missing ordinal):
    $$Z_{ij} \sim \mathcal{N}(\mu_{ij}, \sigma_{ij}^2).$$
  \end{itemize}

  The conditional mean and variance are given by:
  $$Z_{ij} \vert Z_{i,-j}, R \sim \mathcal{N}(\mu_{ij}, \sigma_{ij}^2),$$
  where
  $$\mu_{ij} = R_{j,-j} R_{-j,-j}^{-1} Z_{i,-j}, \quad 
  \sigma_{ij}^2 = R_{jj} - R_{j,-j} R_{-j,-j}^{-1} R_{-j,j}.$$
  
  \item \textbf{Step B: Sample $R$}
  
   According to \cite{hoff2007extending}, the correlation matrix $R$ can be sampled as follows. Given latent variables $\{\mathbf{z}_i\}_{i=1}^n$, sample a covariance matrix:
  $$R^* \sim \mathrm{Inv\text{-}Wishart}
  (\nu_0 + n, \Psi_0 + \sum_{i=1}^n \mathbf{z}_i \mathbf{z}_i^T), $$
  then normalize it to obtain the correlation matrix:
  $$R = \text{diag}(R^*)^{-1/2} R^* \, \text{diag}(R^*)^{-1/2}.$$

  \item \textbf{Step C: Impute missing values in $\bfX$}

  \begin{itemize}
    \item For continuous variables:
    $$ x_{ij} = F_j^{-1}(\Phi(z_{ij})).$$

    \item For ordinal variables:
    $$x_{ij} = \ell \quad \text{if} \quad z_{ij} \in (s_{j,\ell-1}, s_{j,\ell}].$$
  \end{itemize}
\end{itemize}

For each sampled $F \sim P(F \vert \bfX_{\mathcal{O}})$, the BBGC runs multiple Gibbs iterations conditional on $F$, and aggregates the resulting samples to perform Bayesian model averaging over the space of marginal transformations.

Code for the Gibbs sampling procedure is publicly available at \url{https://github.com/zlatjdals/BBGC}.

\section{Simulation Studies}\label{sec:simul_studies}
\subsection{Missing Imputation}
To evaluate the imputation performance of the proposed model, we compared it against various benchmark methods on simulated datasets. The evaluation metric is the normalized root mean squared error (NRMSE) proposed by \cite{oba2003bayesian}:
$$
\mathrm{NRMSE} = \sqrt{\frac{\mathrm{mean}((X^{\text{true}} - X^{\text{imp}})^2)}{\mathrm{Var}(X^{\text{true}})}},
$$
where $X^{\text{true}}$ denotes the true values and $X^{\text{imp}}$ represents the imputed values.

The simulation setup is as follows:
\begin{itemize}
    \item $(n,p) = (1000,15)$,
    \item $R_{ij} = (\lvert i-j \rvert+1)^{-2}$.
\end{itemize}

The data were generated according to the following scheme:
\begin{align}\label{eq:simul}
    (Z_1,\ldots,Z_p) &\overset{iid}{\sim} N(0,R),\\
    X_i &= \lfloor Z_i \rceil, \quad \text{for } i =1,\ldots,5\nonumber\\
    X_i &= F_{U(0,1)}^{-1}\bigl(\Phi(Z_i)\bigr) \quad \text{for } i=6,\ldots,10,\nonumber\\
    X_i &= F_{\text{Exp}(1)}^{-1}\bigl(\Phi(Z_i)\bigr) \quad \text{for } i=11,\ldots,15.\nonumber
\end{align}

Missingness was imposed on the generated dataset $X$ at rates of 10\%, 30\%, 50\%, and 70\% under MCAR and MAR mechanisms. For each missingness level, the imputation performance was evaluated over 100 replications. The competing imputation methods include MICE, KNN, missForest (MF), Dirichlet process mixture model (DPMM), and Mean imputation (MEAN).

\begin{table}[!ht]
\centering
{\scriptsize
\setlength{\tabcolsep}{4pt}
\makebox[\textwidth][c]{ 
\begin{tabular}{lcccccccc}
\toprule
\multirow{2}{*}{Method} 
& \multicolumn{4}{c}{MCAR} 
& \multicolumn{4}{c}{MAR} \\
\cmidrule(lr){2-5} \cmidrule(lr){6-9}
& 10\% & 30\% & 50\% & 70\% & 10\% & 30\% & 50\% & 70\% \\
\midrule
BBGC & \textbf{0.870} (0.013) & \textbf{0.879} (0.008) & \textbf{0.891} (0.006) & 0.914 (0.008) 
     & \textbf{0.870} (0.015) & \textbf{0.879} (0.007) & \textbf{0.890} (0.006) & 0.913 (0.006) \\
MF   & 0.890 (0.014) & 0.917 (0.010) & 0.970 (0.010) & 1.055 (0.015) 
     & 0.890 (0.015) & 0.920 (0.011) & 0.972 (0.012) & 1.057 (0.018) \\
MEAN & 0.903 (0.011) & 0.902 (0.006) & 0.903 (0.005) & \textbf{0.903} (0.005) 
     & 0.901 (0.012) & 0.901 (0.007) & 0.901 (0.005) & \textbf{0.903} (0.005) \\
KNN  & 0.927 (0.012) & 0.929 (0.007) & 0.932 (0.007) & 0.932 (0.007) 
     & 0.925 (0.012) & 0.927 (0.008) & 0.928 (0.007) & 0.929 (0.008) \\
MICE & 1.230 (0.040) & 1.240 (0.025) & 1.254 (0.022) & 1.268 (0.022) 
     & 1.231 (0.035) & 1.244 (0.025) & 1.255 (0.024) & 1.280 (0.024) \\
DPMM & 0.874 (0.013) & 0.883 (0.009) & 0.896 (0.006) & 0.931 (0.011) 
     & 0.873 (0.014) & 0.882 (0.007) & 0.895 (0.007) & 0.932 (0.009) \\
\bottomrule
\end{tabular}
}
}
\caption{Comparison of NRMSE (mean with standard deviation in parentheses) across imputation methods under MCAR and MAR mechanisms on the simulated dataset}
\label{tbl:imp_sim_combined}
\end{table}

Table \ref{tbl:imp_sim_combined} presents the mean and standard deviation of NRMSE for each imputation method under MCAR and MAR settings. For missing rates of 10\%, 30\%, and 50\%, the BBGC method achieves the best performance, exhibiting the lowest NRMSE values among the compared methods. In contrast, at a high missing rate of 70\%, mean imputation achieves the best performance, attaining the lowest NRMSE. These results indicate that while BBGC is highly effective under moderate levels of missingness, mean imputation may provide more stable estimates when the proportion of missing data is extremely high.

\subsection{Uncertainty of Marginal CDF}\
We generated synthetic data $\bfX$ as described in \eqref{eq:simul}, and impose 50\% missingness under the MCAR mechanism to evaluate the uncertainty in the marginal cumulative distribution functions. For this purpose, we select three representative variables: $X_1$, $X_6$, and $X_{11}$, which are known to follow the distributions below:
\begin{align*}
X_1&\overset{d}{=} \lfloor Z\rceil,\quad Z\sim N(0,1)\\
X_6&\sim U(0,1)\\
X_{11}&\sim \text{Exp}(1).
\end{align*}

For each variable, we draw samples of the adjusted Bayesian bootstrap CDF $\tilde{F}^{BB}(t)$ and compute the pointwise 99\% credible interval. Specifically, we collect samples of $\tilde{F}^{BB}(t)$ and calculate the 0.5th and 99.5th percentiles at each $t$ value. These percentiles define the lower and upper bounds of the credible band, which are then connected across the range of $t$ to visualize the uncertainty in the estimated CDF.

\begin{figure}[!ht]
\centering
\includegraphics[width=0.95\linewidth]{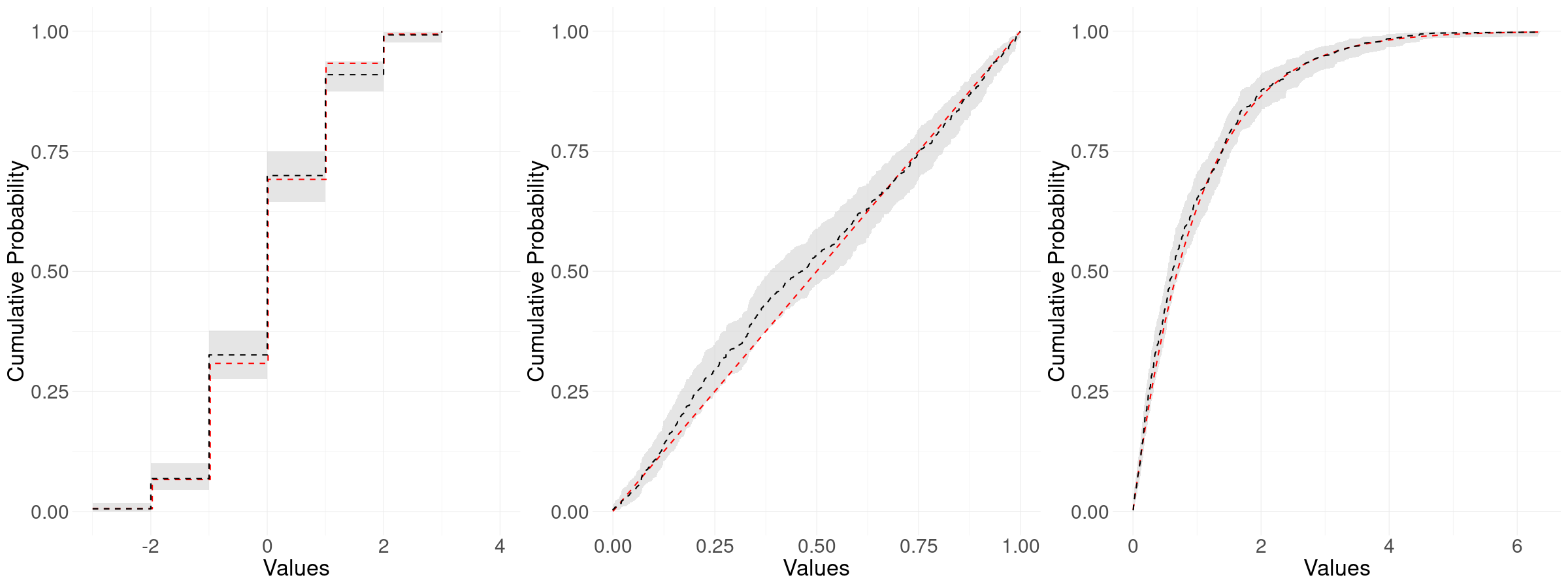}
\caption{Pointwise 99\% credible intervals of marginal cumulative distribution functions for $X_1$, $X_6$, and $X_{11}$ (from left to right).
The shaded gray area indicates the credible region constructed from the adjusted Bayesian bootstrap.
The black dotted line shows the empirical CDF (ECDF) obtained from the observed data, and the red dotted line represents the true CDF of each variable.}
\label{fig:uncert_cdf}
\end{figure}

As shown in Figure~\ref{fig:uncert_cdf}, the 99\% credible intervals (shaded regions) of the adjusted Bayesian bootstrap CDF reliably encompass the empirical CDF (black dotted line) derived from the observed data. Moreover, the true CDF (red dotted line) also lies within the credible region in most cases, indicating that the proposed method successfully captures the uncertainty in the marginal distributions.

For a quantitative assessment, we evaluate the proportion of observed values $x_{ij}$ such that their true CDF values $F_j(x_{ij})$ fall within the 99\% credible interval of $\tilde{F}_j^{BB}(x{ij})$. The resulting coverage rates are 99.2\% for $X_1$, 99.0\% for $X_6$, and 99.8\% for $X_{11}$. These results demonstrate the effectiveness of the adjusted Bayesian bootstrap approach in quantifying uncertainty in the marginal cumulative distribution functions, providing high reliability in practice.

\section{Real Data}\label{sec:real_data}
In this section, we evaluate the performance of imputation methods on three datasets: 
the Breast Cancer Wisconsin (Diagnostic) and Red Wine Quality datasets from the UCI Machine Learning Repository (\citealt{breast_cancer_wisconsin_(diagnostic)_17}; \citealt{wine_quality_186}), 
and semiconductor process data from Samsung Electronics’ DS division.

The Breast Cancer Wisconsin (Diagnostic) dataset includes 569 observations and 31 variables, one of which is binary. The Wine Quality dataset contains 1599 observations and 12 variables, among which three are ordinal. The semiconductor dataset contains 969 observations and 45 variables, with no ordinal variables and a high intrinsic missing rate of 73.2\%.

For the first two datasets, missing values are artificially introduced at rates of 10\%, 30\%, 50\%, and 70\%
under both MCAR and MAR mechanisms. The semiconductor data, which already contains missing values,
is further evaluated under additional MCAR-based scenarios where 50, 100, or 150 entries are randomly removed.
Each experiment is repeated 100 times, and the average normalized root mean squared error (NRMSE) is used as the evaluation metric.

\begin{table}[!ht]
\centering
{\scriptsize
\setlength{\tabcolsep}{4pt}
\makebox[\textwidth][c]{ 
\begin{tabular}{lcccccccc}
\toprule
\multirow{2}{*}{Method} 
& \multicolumn{4}{c}{MCAR} 
& \multicolumn{4}{c}{MAR} \\
\cmidrule(lr){2-5} \cmidrule(lr){6-9}
& 10\% & 30\% & 50\% & 70\% & 10\% & 30\% & 50\% & 70\% \\
\midrule
BBGC & 0.443 (0.027) & 0.486 (0.018) & \textbf{0.528} (0.012) & \textbf{0.573} (0.007)
     & 0.441 (0.027) & 0.478 (0.023) & \textbf{0.519} (0.035) & \textbf{0.563} (0.048) \\
MF   & \textbf{0.340} (0.036) & \textbf{0.454} (0.033) & 0.559 (0.017) & 0.658 (0.022)
     & \textbf{0.344} (0.029) & \textbf{0.445} (0.033) & 0.542 (0.051) & 0.649 (0.068) \\
MEAN & 0.621 (0.022) & 0.622 (0.012) & 0.623 (0.009) & 0.623 (0.006)
     & 0.621 (0.023) & 0.622 (0.018) & 0.622 (0.016) & 0.622 (0.017) \\
KNN  & 0.511 (0.038) & 0.605 (0.016) & 0.628 (0.009) & 0.639 (0.009)
     & 0.508 (0.035) & 0.602 (0.023) & 0.626 (0.024) & 0.626 (0.033) \\
MICE & 0.629 (0.054) & 0.689 (0.031) & 0.743 (0.023) & 0.811 (0.024)
     & 0.622 (0.044) & 0.673 (0.038) & 0.738 (0.052) & 0.807 (0.077) \\
DPMM & 0.443 (0.032) & 0.488 (0.018) & 0.534 (0.014) & 0.579 (0.011)
     & 0.442 (0.032) & 0.482 (0.024) & 0.524 (0.034) & 0.572 (0.045) \\
\bottomrule
\end{tabular}
}
}
\caption{Comparison of NRMSE (mean with standard deviation in parentheses) across imputation methods under MCAR and MAR mechanisms on the Wine Quality dataset}
\label{tbl:imp_real_wine_combined}
\end{table}

Table \ref{tbl:imp_real_wine_combined} summarizes the average and standard deviation of NRMSE for various imputation methods on the winequality dataset across different missing rates. At lower missing rates (10\% and 30\%), the missForest method demonstrates the best performance, whereas at higher missing rates (50\% and 70\%), the BBGC model exhibits superior performance. Furthermore, the Bayesian-based method DPMM achieves NRMSE values that are comparable to those of the BBGC model.

\begin{table}[!ht]
\centering
{\scriptsize
\setlength{\tabcolsep}{4pt}
\makebox[\textwidth][c]{ 
\begin{tabular}{lcccccccc}
\toprule
\multirow{2}{*}{Method} 
& \multicolumn{4}{c}{MCAR} 
& \multicolumn{4}{c}{MAR} \\
\cmidrule(lr){2-5} \cmidrule(lr){6-9}
& 10\% & 30\% & 50\% & 70\% & 10\% & 30\% & 50\% & 70\% \\
\midrule
BBGC & \textbf{0.053} (0.020) & \textbf{0.076} (0.013) & \textbf{0.107} (0.015) & \textbf{0.173} (0.018)
     & \textbf{0.053} (0.022) & \textbf{0.075} (0.014) & \textbf{0.103} (0.013) & \textbf{0.166} (0.027) \\
MF   & 0.129 (0.094) & 0.111 (0.027) & 0.137 (0.017) & 0.202 (0.019)
     & 0.251 (0.032) & 0.248 (0.023) & 0.250 (0.028) & 0.248 (0.027) \\
MEAN & 0.531 (0.035) & 0.536 (0.017) & 0.537 (0.012) & 0.538 (0.008)
     & 0.534 (0.033) & 0.536 (0.017) & 0.539 (0.012) & 0.542 (0.017) \\
KNN  & 0.226 (0.041) & 0.322 (0.023) & 0.420 (0.021) & 0.522 (0.017)
     & 0.238 (0.040) & 0.363 (0.024) & 0.466 (0.022) & 0.527 (0.023) \\
MICE & 0.081 (0.025) & 0.117 (0.016) & 0.175 (0.016) & 0.349 (0.028)
     & 0.085 (0.028) & 0.114 (0.022) & 0.170 (0.019) & 0.331 (0.065) \\
DPMM & 0.148 (0.080) & 0.188 (0.062) & 0.191 (0.056) & 0.188 (0.027)
     & 0.152 (0.089) & 0.181 (0.061) & 0.180 (0.040) & 0.186 (0.032) \\
\bottomrule
\end{tabular}
}
}
\caption{Comparison of NRMSE (mean with standard deviation in parentheses) across imputation methods under MCAR and MAR mechanisms on the Breast Cancer Wisconsin (Diagnostic) dataset}
\label{tbl:imp_real_bc_combined}
\end{table}

Table \ref{tbl:imp_real_bc_combined}summarizes the average and standard deviation of NRMSE for various imputation methods on the Breast Cancer Wisconsin (Diagonostic) dataset across different missing rates. Across all missingness levels, the BBGC model consistently outperforms competing methods, achieving the lowest NRMSE in every scenario.
While MICE performs second-best at a 10\% missing rate, the Bayesian model DPMM surpasses others at the 70\% level, consistently ranking just below BBGC.

\begin{table}[!ht]
\centering
\begin{tabular}{lccc}
\toprule
\multirow[b]{2}{*}{Method} & \multicolumn{3}{c}{Artificial missing count} \\ \cmidrule(lr){2-4}
                          & 50                    & 100                   & 150                  \\
\midrule
BBGC  & \textbf{0.0174} (0.0046)         & \textbf{0.0172} (0.0034)         & \textbf{0.0170} (0.0029)        \\
MF    & 0.0199 (0.0051)         & 0.0213 (0.0037)         & 0.0222 (0.0033)        \\
MEAN  & 0.0199 (0.0052)         & 0.0200 (0.0042)         & 0.0198 (0.0034)        \\
KNN   & 0.0198 (0.0053)         & 0.0199 (0.0041)         & 0.0196 (0.0034)        \\
MICE  & 0.0233 (0.0059)         & 0.0239 (0.0042)         & 0.0236 (0.0032)        \\
DPMM  & 0.0178 (0.0051)         & 0.0178 (0.0035)         & 0.0176 (0.0027)        \\
\bottomrule
\end{tabular}
\caption{Comparison of NRMSE (mean with standard deviation in parentheses) across imputation methods under the MCAR mechanism on the Semiconductor dataset with varying numbers of artificially introduced missing values }
\label{tbl:imp_real_semi}
\end{table}

Table \ref{tbl:imp_real_semi} summarizes the average and standard deviation of NRMSE for various imputation methods across different artificially introduced missing counts in the semiconductor dataset. In all cases, the BBGC method demonstrates the best performance. The Bayesian-based method DPMM consistently ranks as the second-best performer. Notably, since the semiconductor dataset already exhibits a high missing rate of 73.2\%, the performance of alternative methods is comparable to or even worse than that of simple mean imputation.
 
\section{Conclusion}\label{sec:conclusion}
In this paper, we proposed a Bayesian Gaussian Copula model that incorporates the Bayesian bootstrap. By assuming a multivariate normal distribution, the model offers superior interpretability and facilitates both the modeling of inter-variable correlations and prediction through conditional distributions, compared to alternative approaches. Empirical results show that BBGC achieves competitive or superior imputation accuracy, even under high missingness levels across diverse datasets. However, the posterior sampling process in our model is computationally expensive, particularly due to the inversion of the correlation matrix, which poses challenges in high-dimensional settings. To address this, future work may explore structural assumptions on the correlation matrix—such as sparsity or low-rank factorization—to enable faster inversion and improve scalability in high-dimensional scenarios.

\section{Appendix}

\subsection{Proof of Lemma \ref{lem:posterior_mar}}
\begin{proof}
We begin by incorporating the missingness mechanism into the posterior. Let $\mathcal{M} = \mathcal{M}_c\cup \mathcal{M}_o$ denote the missingness indicator matrix, and let $F = (F_1, \ldots, F_p)$ denote the collection of marginal CDFs. Suppose that we are interested in the posterior distribution over $(R, F)$ given the observed data $\bfX_{\mathcal{O}}$ and the missingness pattern $\mathcal{M}$.

The full posterior, including the missingness mechanism, is given by:
\begin{align}
\pi(R, F, \bfX_{\mathcal{M}} \vert \bfX_{\mathcal{O}}, \mathcal{M})
&\propto P(\mathcal{M} \vert \bfX_{\mathcal{O}}, \bfX_{\mathcal{M}}) \cdot \pi(R, F, \bfX_{\mathcal{M}} \vert \bfX_{\mathcal{O}}).
\label{eq:full_joint_missing}
\end{align}

The first term, $P(\mathcal{M} \vert \bfX_{\mathcal{O}}, \bfX_{\mathcal{M}})$, specifies the missing data mechanism—that is, the probability of the missingness pattern $\mathcal{M}$ given the full data $\bfX = (\bfX_{\mathcal{O}}, \bfX_{\mathcal{M}})$. 

Now, under the assumption that the missingness is MAR, we have:
\begin{align}
P(\mathcal{M} \vert \bfX_{\mathcal{O}}, \bfX_{\mathcal{M}}) = P(\mathcal{M} \vert \bfX_{\mathcal{O}}).
\label{eq:mar}
\end{align}

That is, the missingness mechanism depends only on the observed values, and not on the values that are themselves missing. Substituting \eqref{eq:mar} into \eqref{eq:full_joint_missing}, we obtain:
\begin{align*}
\pi(R, F, \bfX_{\mathcal{M}} \vert \bfX_{\mathcal{O}}, \mathcal{M})
&\propto P(\mathcal{M} \vert \bfX_{\mathcal{O}}) \cdot \pi(R, F, \bfX_{\mathcal{M}} \vert \bfX_{\mathcal{O}}).
\end{align*}

Since $P(\mathcal{M} \vert \bfX_{\mathcal{O}})$ is constant with respect to $(R, F, \bfX_{\mathcal{M}})$, it can be dropped from the posterior. Thus, under the MAR assumption, the missingness mechanism is said to be ignorable, meaning that inference for $(R, F)$ can proceed based on the observed data alone:
\begin{align}
\pi(R, F, \bfX_{\mathcal{M}} \vert \bfX_{\mathcal{O}}, \mathcal{M})
= \pi(R, F, \bfX_{\mathcal{M}} \vert \bfX_{\mathcal{O}}).
\label{eq:ignore_M}
\end{align}

It follows that the posterior over $R$ can be obtained by marginalizing over $F$ and the missing values $\bfX_{\mathcal{M}}$:
\begin{align*}
\pi(R \vert \bfX_{\mathcal{O}}, \mathcal{M})
&= \int \int \pi(R, F, \bfX_{\mathcal{M}} \vert \bfX_{\mathcal{O}}, \mathcal{M}) \, d\bfX_{\mathcal{M}} \, dF \\
&= \int \int \pi(R, F, \bfX_{\mathcal{M}} \vert \bfX_{\mathcal{O}}) \, d\bfX_{\mathcal{M}} \, dF.
\end{align*}

To further simplify this expression, we assume that the posterior of $R$ given the full data depends on the observed and missing values only through their transformations into latent Gaussian variables via $F$. Formally, we assume:
\begin{align}
\pi(R \vert \bfX_{\mathcal{O}}, \bfX_{\mathcal{M}}, F) = \pi(R \vert \bfX_{\mathcal{O}}, F).
\label{eq:cond_indep}
\end{align}

This conditional independence is justified by the model structure, where the latent Gaussian variables $Z$ are deterministically related to $\bfX$ through the transformation $F$, and $R$ controls only their dependence.
 
Substituting \eqref{eq:cond_indep} into the marginal posterior, we obtain:
\begin{align*}
\pi(R \vert \bfX_{\mathcal{O}}) 
&\propto \int [ \int \pi(R \vert \bfX_{\mathcal{O}}, F) \cdot \pi(\bfX_{\mathcal{M}} \vert \bfX_{\mathcal{O}}, F) \, d\bfX_{\mathcal{M}} ] \cdot \pi(F \vert \bfX_{\mathcal{O}}) \, dF \\
&= \int \pi(R \vert \bfX_{\mathcal{O}}, F) \cdot \pi(F \vert \bfX_{\mathcal{O}}) \, dF.
\end{align*}
\end{proof}

\subsection{Proof of Theorem \ref{thm:sign_inv_cdf}}
\begin{proof}
    Let $X_{(r)}$ denote the $r$-th smallest value among $X_1, \ldots, X_n$.
    To obtain sign invariance, the following condition must be satisfied:    
    $$
    \bbE_w[\tilde{F}(X_{(r)})] = 1 - \bbE_w[\tilde{F}(X_{(n-r+1)})], \quad \text{for } r = 1, \ldots, n.
    $$
    
    Since $(w_1, \ldots, w_n) \sim \mathrm{Dir}(1, \ldots, 1)$, we have
    \begin{align*}
        \bbE_w[\tilde{F}(X_{(r)})] &= \bbE_w\left[a \sum_{i=1}^r w_i + b \right] \\
        &= a \cdot \frac{r}{n} + b, \\
        \bbE_w[\tilde{F}(X_{(n-r+1)})] &= \bbE_w\left[a \sum_{i=1}^{n-r+1} w_i + b \right] \\
        &= a \cdot \frac{n - r + 1}{n} + b.
    \end{align*}
    
    Therefore, the following equation must hold:
    $$
    a \cdot \frac{r}{n} + b = 1 - \left(a \cdot \frac{n - r + 1}{n} + b \right).
    $$
    
    Simplifying, we obtain the condition:
    $$
    \frac{n + 1}{n}a + 2b = 1.
    $$
\end{proof}

\subsection{Proof of Theorem \ref{thm:post_conv_cdf}}
\begin{proof}
    Let $D_n = \sup_{t \in \mathbb{R}} \abs{F_n(t) - F(t)}$ and  
    $C_n = \sup_{t \in \mathbb{R}} \abs{F^{BB}(t) - F(t)}$,  
    where $F_n$ is the empirical cumulative distribution function (ECDF) and $F^{BB}$ is the Bayesian bootstrap CDF.
    
    Given that $U_i = F(X_i) \overset{iid}{\sim} U(0,1)$ for $i = 1, \ldots, n$, we can rewrite $C_n$ as follows:
    $$
    \begin{aligned}
    C_n &= \sup_{t \in \mathbb{R}} \abs{F^{BB}(t) - F(t)} \\
    &= \sup_{t \in \mathbb{R}} \abs{\sum_{i=1}^n w_i \mathbf{1}(X_i \leq t) - F(t)} \\
    &= \sup_{t \in \mathbb{R}} \abs{\sum_{i=1}^n w_i \mathbf{1}(F(X_i) \leq F(t)) - F(t)} \\
    &= \sup_{F(t) \in [0,1]} \abs{\sum_{i=1}^n w_i \mathbf{1}(U_i \leq F(t)) - F(t)} \\
    &= \sup_{s \in [0,1]} \abs{\sum_{i=1}^n w_i \mathbf{1}(U_i \leq s) - s}.
    \end{aligned}
    $$
    
    The supremum over $s \in [0,1]$ can be computed via order statistics of the uniform variables. Let $U_{(k)}$ denote the $k$-th order statistic of $\{U_1, \ldots, U_n\}$. Then,
    $$
    C_n = \max_{1 \leq k \leq n} [ \max ( | \sum_{i=1}^k w_i - U_{(k)} |, | \sum_{i=1}^{k-1} w_i - U_{(k)} | ) ].
    $$
    
    Similarly, we can express $D_n$ as:
    $$
    D_n = \max_{1 \leq k \leq n} [ \max ( | \frac{k}{n} - U_{(k)} |, | \frac{k-1}{n} - U_{(k)} | ) ].
    $$
    
    Since $(w_1, \ldots, w_n) \sim \mathrm{Dirichlet}(1,\ldots,1)$, the partial sums of weights follow the distribution of sorted uniform random variables:
    $$
    (w_1, \sum_{i=1}^2 w_i, \ldots, \sum_{i=1}^{n-1} w_i ) \overset{d}{=} (V_{(1)}, \ldots, V_{(n-1)}),
    $$
    where $V_i \overset{iid}{\sim} U(0,1)$ and $V_{(k)}$ is the $k$-th order statistic.
    
    By the Dvoretzky-Kiefer-Wolfowitz (DKW) inequality with Massart’s sharp constant (\cite{massart1990tight}), we have:
    $$
    \mathbb{P}(D_n > \epsilon) \leq 2 \exp(-2n \epsilon^2), \quad \text{for all } \epsilon > 0.
    $$
    
    Applying the same logic to the Dirichlet weights, define
    $$
    E_{n-1} = \max_{1 \leq k \leq n-1} [ \max ( | \frac{k}{n-1} - V_{(k)} |, | \frac{k-1}{n-1} - V_{(k)} | ) ],
    $$
    and again by the DKW inequality,
    $$
    \mathbb{P}(E_{n-1} > \epsilon) \leq 2 \exp(-2(n-1)\epsilon^2).
    $$
    
    Now consider the difference between $U_{(k)}$ and $V_{(k)}$. For each $k$,
    $$
    \begin{aligned}
    | V_{(k)} - U_{(k)} | 
    &\leq | V_{(k)} - \frac{k}{n-1} | + | \frac{k}{n} - U_{(k)} | + \frac{k}{n(n-1)}, \\
    | U_{(k)} - V_{(k-1)} | 
    &\leq | U_{(k)} - \frac{k-1}{n} | + | \frac{k-1}{n-1} - V_{(k-1)} | + \frac{k-1}{n(n-1)}.
    \end{aligned}
    $$
    
    Hence,
    $$
    C_n \leq D_n + E_{n-1} + \frac{1}{n-1}.
    $$
    
    Therefore, for any $\delta > \dfrac{1}{n-1}$,
    $$
    \begin{aligned}
    \pi( \sup_{t \in \mathbb{R}} \abs{F^{BB}(t) - F(t)} > \delta \vert \bfX )
    &= \pi(C_n > \delta \vert \bfX) \\
    &\leq \mathbb{P}( D_n + E_{n-1} + \frac{1}{n-1} > \delta ) \\
    &\leq \mathbb{P}( D_n > \frac{1}{2}(\delta - \frac{1}{n-1} ) ) + \mathbb{P}( E_{n-1} > \frac{1}{2}(\delta - \frac{1}{n-1} ) ) \\
    &\leq 2 \exp( -2n ( \frac{1}{2}(\delta - \frac{1}{n-1} ) )^2 ) \\
    &\quad + 2 \exp( -2(n-1) ( \frac{1}{2}(\delta - \frac{1}{n-1} ) )^2 ) \\
    &\leq 4 \exp( -\frac{1}{2}(n-1) ( \delta - \frac{1}{n-1}  )^2 ).
    \end{aligned}
    $$

    Finally, since $\tilde{F}^{BB}(t) = \dfrac{n}{n+1}F^{BB}(t)$, the inequality holds
    $$\sup_{t\in\bbR}\abs{\tilde{F}^{BB}(t)-F^{BB}(t)}\leq \dfrac{1}{n+1}.$$

    Therefore, for any $\delta>\dfrac{2n}{n^2-1}$,
    \begin{align*}
        \pi( \sup_{t \in \mathbb{R}} \abs{\tilde{F}^{BB}(t) - F(t)} > \delta \vert \bfX ) &\leq \pi( \sup_{t \in \mathbb{R}} \abs{F^{BB}(t) - F(t)} + \sup_{t \in \mathbb{R}} \abs{\tilde{F}^{BB}(t) - F^{BB}(t)} > \delta \vert \bfX )\\
        &\leq  \pi( \sup_{t \in \mathbb{R}} \abs{F^{BB}(t) - F(t)}  > \delta  - \dfrac{1}{n+1}\vert \bfX ) \\
        &\leq 4 \exp( -\dfrac{1}{2}(n-1) ( \delta - \frac{2n}{n^2-1}  )^2 ).
    \end{align*}
    
\end{proof}

\section*{Acknowledgements}
This research was supported in part by Samsung Electronics.

\bibliographystyle{dcu}
\bibliography{copula}

\end{document}